# Remarks to the Article: “New Light on the Invention of the Achromatic Telescope Objective”

Igor Nesterenko
*FRIB/NSCL, Michigan State University, East Lansing, MI 48824, USA*
*Budker Institute of Nuclear Physics, Novosibirsk, 630090, RUSSIA*
(corresponding author, e-mail: nesterenko@frib.msu.edu)

## Abstract
The article analysis was carried out within the confines of the replication project of the telescope, which was used by Mikhail Lomonosov at observation the transit of Venus in 1761. At that time he discovered the Venusian atmosphere. It is known that Lomonosov used Dollond’s 4.5 feet long achromatic telescope. The investigation revealed significant faults in the description of the approximation method, which most likely was used by J. Dollond & Son during manufacturing of the early achromatic lenses.

## Introduction
In the article [1] R. Willach described the research of the four early achromatic lenses. Two doublet lenses were made by Dollond: one is flint-forward type and other lens is crown-forward type. The others two doublet lenses (both flint-forward type) were made by James Ayscough and by Joseph Linnell[1]. The flint-forward doublets are classified as a first early achromatic lenses in comparison with crown-forward type.
The optical parameters of the examined flint-forward doublets are collected in the Table 1. For calculation and comparison of the optical systems, Zemax-EE software was used.

Table 1. Optical parameters of the flint-forward achromatic lenses

| Maker Name | D (mm) | F (mm) | $R_1$ (mm) | $R_2$ (mm) | $t_{flint}$ (mm) | Glass Name | $t_{air}$ (mm) | $R_3$ (mm) | $R_4$ (mm) | $t_{crown}$ (mm) | Glass Name |
|---|---|---|---|---|---|---|---|---|---|---|---|
| Dollond | 32.0 | 773 | -1826 | 190 | 2.8 | E18_F1 | 0.0 | 193 | -262 | 3.7 | E18_C1 |
| Linnell | 24.5 | 492 | -83000 | 99 | 1.9 | E18_F2 | 0.3 | 136 | -132 | 4.0 | E18_C2 |
| Ayscough | 32.0 | 790 | -803 | 168 | 1.1 | E18_F3 | 0.2 | 220 | -171 | 3.5 | E18_C3 |

The measured refractive indexes of the glasses from the investigated achromatic doublets at the different wavelengths are described in Table 2. These refractive indexes were used for finding the dispersion dependencies.

Table 2. Refractive indexes of the glasses.

| Glass Name | Refraction Index (wavelength, μm) | | | Abbe Number[2] | |
|---|---|---|---|---|---|
| | $n_F$(0.48613) | $n_e$(0.54607) | $n_C$(0.65628) | $\nu_e$(0.54607) | $\nu_d$(0.58756) |
| E18_F1 | 1.5930 | 1.5870 | 1.5803 | 46.22 | 45.99 |
| E18_F2 | 1.5822 | 1.5759 | 1.5688 | 42.98 | 42.74 |
| E18_F3 | 1.5873 | 1.5810 | 1.5747 | 46.11 | 45.88 |
| E18_C1 | 1.5264 | 1.5227 | 1.5185 | 66.16 | 65.93 |
| E18_C2 | 1.5285 | 1.5246 | 1.5199 | 61.00 | 60.78 |
| E18_C3 | 1.5272 | 1.5230 | 1.5187 | 61.53 | 61.30 |

[1] He was free of the Spectacle Makers Company (SMC) on June 30, 1763 and he took over the business from hands of Ayscough widow in 1764. Maybe J. Linnell was apprenticed to J. Ayscough sometime between 1754 and 1759.
[2] $\nu(\lambda) = [n(\lambda) - 1]/[n(0.48613) - n(0.65628)]$, λ – wavelength in μm at which is found Abbe number.

For fitting of the dispersion dependencies the Conrady formula was used:

$$n(\lambda) = n_0 + {}^{A}/_{\lambda} + {}^{B}/_{\lambda^{3.5}}$$

Here $n_o$, $A$, $B$ – free parameters, which can be found by using the three measured refractive indexes at the different wavelengths. These Conrady parameters for each glass are presented in Table 3, it is assumed that wavelength is expressed in µm.

Table 3. Conrady parameters of the glasses

| Glass Name | Conrady Parameters | | |
|---|---|---|---|
| | $n_o$ | A | B |
| E18_F1 | 1.55993 | $1.0948\cdot10^{-2}$ | $8.4524\cdot10^{-4}$ |
| E18_F2 | 1.54652 | $1.2182\cdot10^{-2}$ | $8.5062\cdot10^{-4}$ |
| E18_F3 | 1.56335 | $3.6944\cdot10^{-3}$ | $1.3096\cdot10^{-3}$ |
| E18_C1 | 1.50500 | $7.4741\cdot10^{-3}$ | $4.8217\cdot10^{-4}$ |
| E18_C2 | 1.50196 | $1.0766\cdot10^{-2}$ | $3.5181\cdot10^{-4}$ |
| E18_C3 | 1.50977 | $3.5247\cdot10^{-3}$ | $8.1560\cdot10^{-4}$ |

For the described optical systems on Figure 1, the following focal lengths at wavelength 0.546µm were received: for Dollond's doublet 771.1mm instead of measured 773mm;

for Linnell's doublet 490.1mm (was measured 492mm);

for Ayscough's doublet 785.7mm (was measured 790mm).

The observable differences can be explained by existence of additional spherical aberration due to imperfection of the real surfaces and by precision of the measurement method of the focal length.

**Dollond**

| Surf | Type | Comment | Radius | Thickness | Glass | Semi-Diameter | Conic |
|---|---|---|---|---|---|---|---|
| OBJ | Standard | | Infinity | Infinity | | 0.000 | 0.000 |
| STO | Standard | | Infinity | 0.250 | | 16.100 U | 0.000 |
| 2* | Standard | R1 | -1826.000 | 2.800 | E18_F1 | 18.000 U | 0.000 |
| 3* | Standard | R2 | 190.000 | 0.014 | | 18.000 U | 0.000 |
| 4* | Standard | R3 | 193.000 | 3.700 | E18_C1 | 18.000 U | 0.000 |
| 5* | Standard | R4 | -262.000 | 771.921 M | | 18.000 U | 0.000 |
| IMA | Standard | | Infinity | - | | 0.033 | 0.000 |

**Linnell**

| Surf | Type | Comment | Radius | Thickness | Glass | Semi-Diameter | Conic |
|---|---|---|---|---|---|---|---|
| OBJ | Standard | | Infinity | Infinity | | 0.000 | 0.000 |
| STO | Standard | | Infinity | 0.250 | | 12.300 U | 0.000 |
| 2* | Standard | R1 | -8.300E+004 | 1.900 | E18_F2 | 14.000 U | 0.000 |
| 3* | Standard | R2 | 99.000 | 0.273 | | 14.000 U | 0.000 |
| 4* | Standard | R3 | 136.000 | 4.000 | E18_C2 | 14.000 U | 0.000 |
| 5* | Standard | R4 | -132.000 | 496.550 M | | 14.000 U | 0.000 |
| IMA | Standard | | Infinity | - | | 0.118 | 0.000 |

**Ayscough**

| Surf | Type | Comment | Radius | Thickness | Glass | Semi-Diameter | Conic |
|---|---|---|---|---|---|---|---|
| OBJ | Standard | | Infinity | Infinity | | 0.000 | 0.000 |
| STO | Standard | | Infinity | 0.250 | | 16.100 U | 0.000 |
| 2* | Standard | R1 | -803.000 | 1.100 | E18_F3 | 18.000 U | 0.000 |
| 3* | Standard | R2 | 168.000 | 0.230 | | 18.000 U | 0.000 |
| 4* | Standard | R3 | 220.000 | 3.500 | E18_C3 | 18.000 U | 0.000 |
| 5* | Standard | R4 | -171.000 | 791.537 M | | 18.000 U | 0.000 |
| IMA | Standard | | Infinity | - | | 0.053 | 0.000 |

Figure 1. The optical systems of the achromatic lenses with flint-forward design. The air gaps between flint and crown lenses were selected for providing mechanical contacts on the lens edges.

Figures 2 and 3 show chromatic and geometric aberrations of the three achromatic doublets. The chromatic compensations in all lenses are not ideal. The standard for achromatic lens the parabolic dependence for residual chromatic focal shift was not reached. Nevertheless, Ayscough's and Dollond's doublets with Strehl ratio about 0.9 can be classified as diffraction limited (DL) optics, mainly due to small enough relative apertures *A* (1/24.6 and 1/24.1 respectively). Linnell's doublet has a big spherical aberration, which is greater than the diffraction limit mainly due to its bigger relative aperture (1/20) in comparison with the previous two doublets.

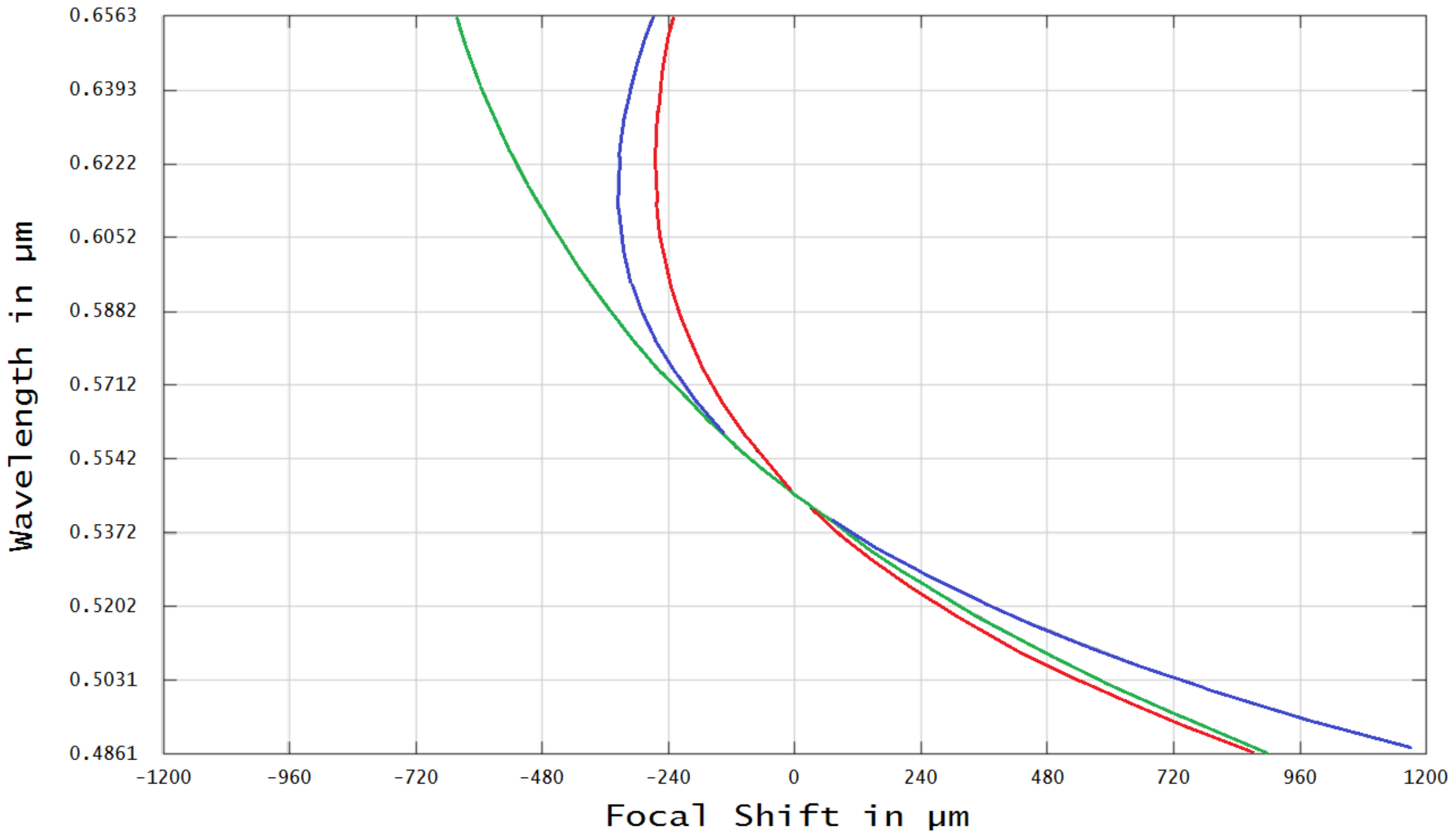

Figure 2. The residual chromatic focal shift dependencies of the achromatic lenses. The red solid line is the focal shift for Ayscough's achromatic lens, the blue solid line is for Linnell's lens and the green solid line is for Dollond's lens.

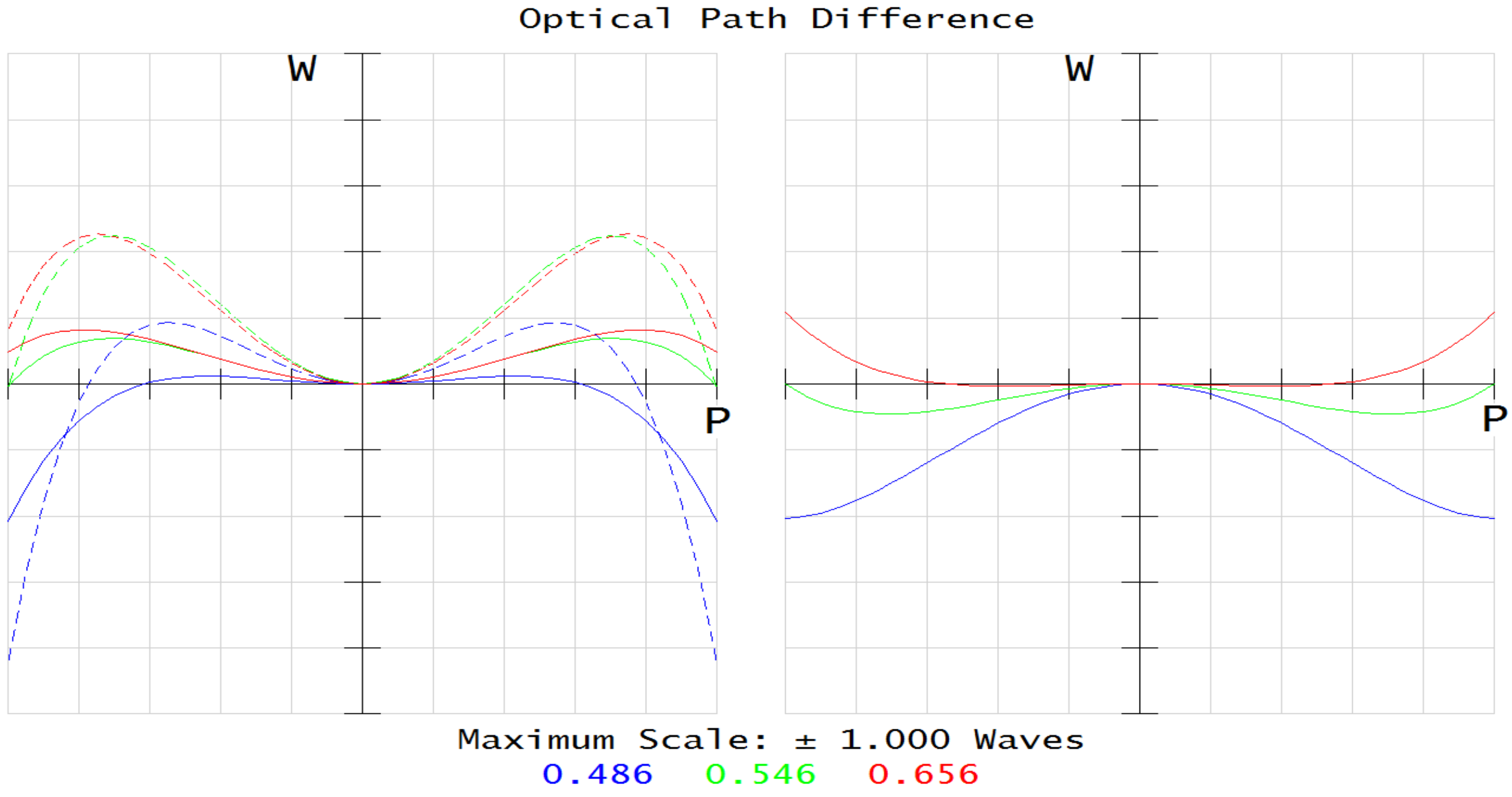

Figure 3. The optical path difference (OPD) versus of the pupil coordinate at different wavelength (0.486μm, 0.546μm and 0.656μm). On the left picture the solid lines are for Ayscough's lens and the dashed lines for Linnell's lens. Both doublets have an over-corrected spherical aberration. On the right picture are shown OPD dependencies for Dollond's lens with under-corrected spherical aberration.

Here it should be reminded that a transverse spherical aberration is proportional to $A^3$ and an axial chromatic (transverse) aberration is proportional to $A$. Therefore any approximate calculation methods of the primary parameters for the achromatic doublets can be properly comparable to each other only at equal relative apertures. Consequently, if the OPD dependencies[3] for Linnell's doublet (dashed lines) on Figure 3 will be divided on factor $(24.6/20)^4 \approx 2.3$, then these curves will be very close to results for Ayscough's lens.

**Initial approximations**

At preliminary calculation of a doublet system in "thin lens" approximation should be specified four radius. Such optical system is described by two conditions. These are the selected focal length $F$ (or total optical power $P=1/F$) and that it is the achromatic doublet for the selected glass pair. Hence two radii continue to be free parameters. Herewith the versions of the optical systems should be chosen with a defocusing flint lens ($P_f < 0$) and a focusing crown lens ($P_c > 0$), because the flint glass has bigger dispersion in comparison with crown glass, and also the total optical power of the achromatic doublet has to have positive value.
Possible versions[4] of the flint-forward doublets are shown on Figure 4. Versions #7, #10 and #11 have the evident technical issue: the crown lens rests on the flint lens in one vertex point and without supporting by the spacer ring between the lenses it will be tilted in during rolling the back rim of a lens mount.

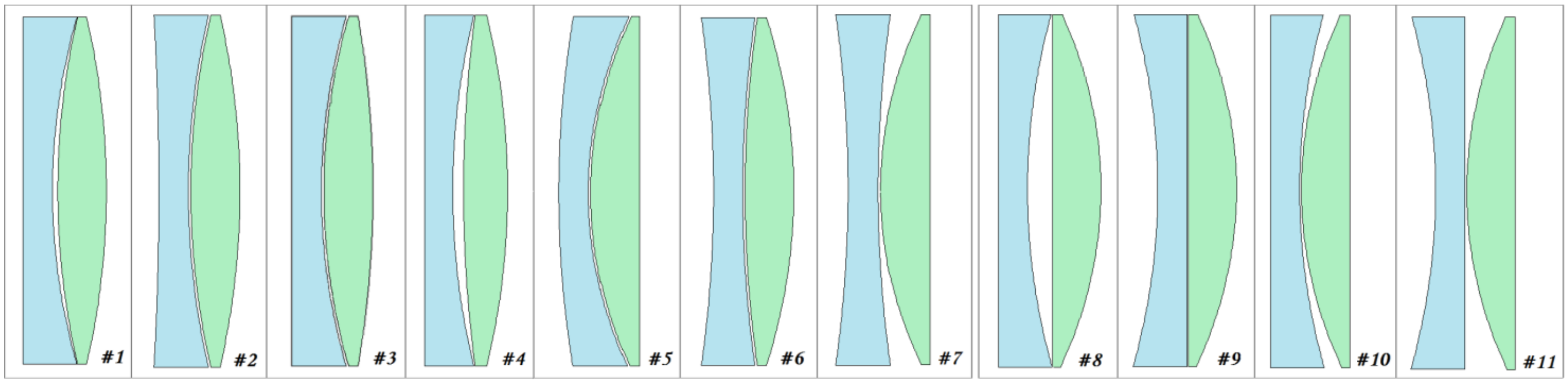

Figure 4. Some approximation versions of the flint-forward achromatic doublets are shown. The blue part is the flint lens, light go from left to right. Version #4 is different from #3th only by reverse orientation of the crown lens, all the radii are the same.

A comparison of these versions was done for the optical systems with aperture 32mm and focal length 640mm (focal ratio 20) for glass pair: flint E18_F2 ($n_e$=1.5759, $\nu_e$=42.98) and crown E18_C2 ($n_e$=1.5246, $\nu_e$=61.00). The calculation results are collected in Table 4.
The last four versions beginning from #8 were reported by William Eastland in 1765 during the testimony in the court. He referred to the instruction for making an achromatic lens by James Ayscough, who wrote down it from words of Chester Moor Hall [2]. In the court records it was written that both flint and crown lenses have to have a flat surface on one of sides. The other sides of the lenses should be concave for flint and convex for crown. Their order and relative orientation were not specified. Therefore all four possible the flint-forward versions were included in the consideration.
Relying on the results from Table 4, we can definitely say that from the last four versions (authorship attributed to Chester Moor Hall) only one (#8) may be used as initial approximation for achromatic doublets. The other versions are so far away from the optimum, mainly due to big original spherical aberration, therefore they hardly were used in practice. At least, their application would hardly led to successful results. The same conclusion can be applied in respect to versions #6 and #7.
Versions #2, #4 and #8 may be used in case if enough accurate knowledge about the glass parameters are present and the calculated radii are kept within tolerance, and that obligatory subsequent correction of a spherical aberration will be finally done. Otherwise many steps of the trial-and-error method will be required and as results loss of producing efficiency. Other possible way, it is reduction of the relative aperture for the designed achromatic lens that it is desirable for these approximation versions.

---

[3] For pure spherical aberration OPD (or deformation of the ideal wave front) is proportional to $A^4$.
[4] In really the total number of the possible versions equal to 20. More details can be found in Appendix I.

Table 4. Comparison of the versions of the initial approximations for the flint-forward achromatic doublets[5]

| # | $R_1$ | $R_2$ | $R_3$ | $R_4$ | F/D | Peak-to-Valley (λ=0.546μm) | Strehl Ratio | GEO Radius (μm) | RMS Radius (μm) | TSPH (μm) | TAXC (μm) |
|---|---|---|---|---|---|---|---|---|---|---|---|
| 1 | ∞ | $+R_2$ | +R | −R | 19.9 | 0.185λ | 0.887 | 33.0 | 13.5 | -60.8 | 2.9 |
| 2 | $-R_1$ | +R | +R | −R | 19.9 | 0.297λ | 0.734 | 51.7 | 21.4 | 102.7 | 1.9 |
| 3 | ∞ | +R | +R | $-R_4$ | 20.0 | 0.116λ | 0.954 | 20.2 | 8.4 | 40.5 | 0.0 |
| 4 | ∞ | +R | $+R_3$ | −R | 19.7 | 0.290λ | 0.749 | 51.5 | 21.1 | -93.5 | 6.9 |
| 5 | $+R_1$ | +R | +R | ∞ | 20.0 | 0.069λ | 0.983 | 12.2 | 5.0 | -23.6 | 1.5 |
| 6 | −R | +R | +R | $-R_4$ | 19.7 | 0.622λ | 0.258 | 107.2 | 44.3 | 212.0 | 6.2 |
| 7 | −R | +R | $+R_3$ | ∞ | 19.9 | 2.741λ | ~0.04 | 458.3 | 188.6 | 904.5 | 0.2 |
| 8 | ∞ | $+R_2$ | ∞ | $-R_4$ | 19.2 | 0.350λ | 0.655 | 57.8 | 24.3 | 125.6 | 18.9 |
| 9 | $-R_1$ | ∞ | ∞ | $-R_4$ | 19.1 | 1.534λ | ~0.10 | 252.7 | 104.1 | 493.8 | 15.9 |
| 10 | ∞ | $+R_2$ | $+R_3$ | ∞ | 20.0 | 1.785λ | ~0.06 | 305.7 | 125.7 | 594.5 | -1.1 |
| 11 | $-R_1$ | ∞ | $+R_3$ | ∞ | 19.7 | 3.033λ | ~0.04 | 498.9 | 205.0 | 980.2 | 2.7 |

Thus, only three versions #1, #3 and #5 can be confidently applied at the focal ratio 20 as the initial approximation for the achromatic doublets.
More details about calculations of the radii for the described versions can be found in Appendix I.
On the next step we have to answer on the question:
Which from the versions most likely were used by Chester Moor Hall and by company J. Dollond & Son for calculation and making the achromatic lenses?

**Versions selection**

Firstly, only six initial approximations from the Table 4 will be used. These are the versions from #1 to #5 and #8. Other versions are excluded because they give a bad approximation with a big original spherical aberration, on which we will have to spend efforts to eliminate. Quality of the chromatic correction is defined foremost by knowledge of the glass parameters or by measurement accuracy theirs. Herewith, all possible flint-forward versions have comparable quality of a chromatic corrections, and residual chromatic aberrations are much smaller in comparison to the spherical aberrations.
Secondly, it is necessary to use imperial length unit (inch) instead of the metric unit (mm). Because at the time, when these three investigating telescopes were made, the metric system was absent. In other words the same length unit system should be used which was used by the telescope makers.
Thirdly, at the selection should be given away priority to the grinding tools with a radii of integer value in applied the system of length unit. Radii with fractional part of the length unit (half-integer and etc.) should be evaluated in descending order in probability of their application. It's easier to slightly change the selected focal length to get an integer or half-integral radius than to use the radius with a small fractional part. In addition, at a large spread the parameters of flint glass, it is easier to slightly vary the focal length of achromatic lens under the existing set of a grinding tools.
Finally, as a quantitative criterion for validity of the particular version will be used a proximity of the calculated radii to the real measured radii from Table 1. In other words, the version validity will be described in terms of the differences in the sags between the measured and calculated surfaces. Main reason why the surface sags were selected for forming the quantitative criterion – these are directly measurable values in contrast to the radii.
By analogy with $\chi^2$ distribution, which is widely used for testing applicability of a statistical hypothesis, will be used the next functional $\gamma^2$ for comparison of different approximation versions:

$$\gamma^2 = \sum_{i=1}^{4} \left(\frac{z_{mi} - z_{ci}}{\delta}\right)^2$$

[5] Here in the table:
$R_1$, $R_2$ – first and second radii (by ray path) of the flint lens; $R_3$, $R_4$ – the similar radii for the crown lens. The signs before the radii are selected in accordance with modern rules applying at description of the optical systems;
F/D – the focal ratio obtained in Zemax's calculations with considering a lens thickness and required air gap glasses;
GEO, RMS – the envelope and RMS radii respectively on the spot diagram (for reference, the Airy radius is about 13μm);
TSPH, TAXC –spherical and axial chromatic aberration coefficients respectively, letter 'T' means the transverse. The negative sign for the spherical and for the axial chromatic aberrations indicates about over-correction.
Peak-to-Valley and Strehl Ratio are calculated for the focal plane, where is minimum RMS of wave front error.

Here $z_{mi}$ and $z_{ci}$ – the measured and calculated sags[6] respectively for i-th surface; δ - accuracy[7] of the sag measurements. From the formula it is easy to see than the larger $\gamma^2$ value (more deviation) the less validity of the particular version.

*Dollond's doublet*

Aperture is 32mm that is slightly over 1.25" and focal length is 773mm or almost 30.5".
The measured radii of the lenses after recalculation to Imperial system are
$R_1 \approx -71.89$" (-1826mm); $R_2 \approx 7.48$" (190mm); $R_3 \approx 7.60$" (193mm) and $R_4 \approx -10.31$" (-262mm);
For the six selected versions the calculated radii with rounding to quarter inch at the focal length 29.5" (749mm) are collected in the table below:

| | $R_1$ | $R_2$ | $R_3$ | $R_4$ | $\gamma^2$ | F(mm) | SR[8] |
|---|---|---|---|---|---|---|---|
| #1 | ∞ | 7.5" | 9.25" | -9.25" | 22240 | 727 | 0.980 |
| #2 | -38" | 9.25" | 9.25" | -9.25" | 37990 | 741 | 0.890 |
| #3 | ∞ | 7.5" | 7.5" | -12.25" | 10980 | 744 | 0.971 |
| #4 | ∞ | 7.5" | 12.25" | -7.5" | ~103000 | 740 | 0.942 |
| #5 | 12.25" | 4.75" | 4.75" | ∞ | ~785000 | 738 | 0.965 |
| #8 | ∞ | 7.5" | ∞ | -4.75" | ~779000 | 772 | 0.876 |

The focal lengths (F) and Strehl ratios (SR) in the table were calculated taking into account the measured thickness of the lenses and required gaps between them in the doublet.
All versions show the diffraction limited quality, which is not surprising because the relative aperture smaller 1/22.7 and transverse spherical aberration should be smaller about 1.5 times than previously for Table 4.
According to the $\gamma^2$ values only version #3 has an explicit preference or it has better validity as the initial approximation to the real doublet.
However, it should be said that information about the selectable focal length, which was used by the maker for the preliminary calculation is absent. Therefore, it is necessary to test the versions also at other focal length values. For example, if to take for version #2 the focal length 23.8" (604.52mm), then the radii will be
$R_1 \approx -30.75$" (-781.05mm); $R_2 = R_3 = -R_4 \approx 7.5$" (190.5mm).
For this variant $\gamma^2$ is about 42770. It's worse than the previous result from the table. Nevertheless, $\gamma^2$ minimum exists and it is located near next radii:
$R_1 \approx -34.75$" (-882.65mm); $R_2 = R_3 = -R_4 \approx 8.5$" (215.9mm).
This radii set is obtained at the focal length 26.9" (683mm). In this case $\gamma^2$ is about 28160 and version #2 still is worse or it has smaller validity in comparison with version #3.
If to vary the radii for version #1 with a view of searching $\gamma^2$ minimum it will be found near the radii:
$R_1 = \infty$; $R_2 \approx 7.25$" (184.15mm); $R_3 = -R_4 \approx 9.0$" (228.6mm);

---

[6] The sag is calculated by formula: $z = \frac{D^2}{4R\left(1+\sqrt{1-\left(D/_{2R}\right)^2}\right)}$, here D – aperture of the achromatic lens and R – radius of the surface.

[7] With a high probability at sag measurements R. Willach used a micrometer head with resolution 1μm, because in Linnell's lens first radius -83000mm has a sag on the aperture 24.5mm about 0.001mm and it was distinguished from a flat surface.

[8] In this case Strehl Ratio was calculated for three wavelength 0.4861μm, 0.5461μm and 0.6563μm in sum with weight coefficients 0.17, 1.0 and 0.1 respectively.

which correspond to the focal length about 28.6” (726mm). For this case $\gamma^2$ is about 21270 that again worse than for version #3 about two times.

Here arises the question, why it was necessary to introduce the changes into the initial approximation (version #3), which immediately gives very good result (Strehl ratio 0.97) at this relative aperture?

Apparently it was decided that the original flat surface on the flint lens has a bad optical quality, especially after polishing the second surface. It is well-known that the opposite surface will be twisted after the material removal. Therefore, it is quite possible from the set of grinding tools existing in the optical workshop the one of the largest radius was chosen. In case of this doublet the tool was for concave radius -72”. As a result the power of the flint lens was increased. After that the optician was forced to change one of the surfaces on the crown lens to increase of its power in the known proportion for saving of the achromatic condition. The easiest way is to re-polish the external surface on the crown lens.
The ratio $v_c / v_f$ from the achromatic condition is about 1.431 for E18_C1 and E18_F1 glass pair, which is used in this doublet. For version #3 the radii from the table give ratio $P_c / |P_f|$ about 1.436. After changing of the first radius from flat to radius -72”, the optician have to select the grinding tool for the convex radius on the fourth surface between 9.5” (241.3mm) and 10” (257mm). After that ratio $P_c / |P_f|$ will be returned back in a range from 1.411 to 1.443.
For “re-polished” version #3 the new radii, which save achromaticity of the doublet, will be
$R_1 \approx -72$” (1828.8mm); $R_2 = R_3 \approx 7.5$” (190.5mm); $R_4 \approx -9.75$” (247.65mm); and $\gamma^2$ drops down to 890.
Similar re-polishing manipulations give the new radii
for version #1: $R_1 \approx -72$”; $R_2 \approx 7.5$”; $R_3 \approx 9.25$” (234.95mm); $R_4 \approx -7.75$” (247.65mm); and $\gamma^2 \approx 40490$;
for version #2: $R_1 \approx -72$”; $R_2 = R_3 \approx 9.25$”; $R_4 \approx -11.5$” (292.1mm); and $\gamma^2 \approx 33450$.
The differences in the validity of the versions became even more evident. That is an additional argument in favor of version #3.
All these reasoning and recalculations could be done easily before of making the real lenses. Evidently that optician was polishing the lenses with the required radii right away relying on the preliminary calculations.

*Linnell’s doublet*

Aperture is 24.5mm that hardly less than 1” and focal length is 492mm or almost 19.5”.
The measured radii of the lenses after recalculation to Imperial system are
$R_1 \approx -3268$” (-83007mm); $R_2 \approx 3.90$” (99mm); $R_3 \approx 5.35$” (136mm) and $R_4 \approx -5.19$” (-132mm).
For the selected versions the calculated radii with rounding to quarter inch are collected in the table below:

| | $R_1$ | $R_2$ | $R_3$ | $R_4$ | $\gamma^2$ | F(mm) | SR |
|---|---|---|---|---|---|---|---|
| #1 | ∞ | 4” | 5.25” | -5.25” | 535 | 449 | 0.659 |
| #1 | ∞ | 4” | 5” | -5” | 2440 | 382 | 0.921 |
| #2 | -17.75” | 5” | 5” | -5” | 57900 | 404 | 0.539 |
| #3 | ∞ | 4” | 4” | -7” | 57520 | 408 | 0.878 |
| #4 | ∞ | 4” | 7” | -4” | 46800 | 399 | 0.693 |
| #5 | 7” | 2.5” | 2.5” | ∞ | ~1100000 | 411 | 0.913 |
| #8 | ∞ | 4” | ∞ | -2.5” | ~694320 | 362 | ~0.258 |

First line in the table corresponds to selectable focal length 16.9” (429mm). Second and all rows below correspond to the case, if the optician selected the focal length 16.2” (411mm).
The results for $\gamma^2$ definitely indicate on the version #1. Undoubtedly the first variant is closer to the real doublet, but it has a larger spherical aberration (at relatively small and comparable chromatic aberration) TSPH=−91.4µm instead of −27.1µm for the second variant. It is due to the rounding of the radii to quarter inch. The observable dispersion in quality of the initial approximation can be interpreted as a stability of the selected version to any errors at relative aperture of about 1/16. On the other hand, the rounding may be interpreted also as a skill to keep the radii with specified tolerance at making of a lenses.

*Ayscough's doublet*

Aperture is 32mm that is slightly over 1.25" and focal length is 790mm or almost 31".
The measured radii of the lenses after recalculation to Imperial system are
$R_1 \approx -31.61$" (-803mm); $R_2 \approx 6.61$" (168mm); $R_3 \approx 8.66$" (220mm) and $R_4 \approx -6.73$" (-171mm);
For the selected versions the calculated radii with rounding to quarter inch at the focal length 33.3" (846mm) are collected in the table below:

| | $R_1$ | $R_2$ | $R_3$ | $R_4$ | $\gamma^2$ | F(mm) | SR |
|---|---|---|---|---|---|---|---|
| #1 | ∞ | 6.5" | 8.75" | -8.75" | 55740 | 835 | 0.742 |
| #2 | -25" | 8.75" | 8.75" | -8.75" | 66880 | 844 | 0.896 |
| #3 | ∞ | 6.5" | 6.5" | -13.5" | 205330 | 851 | 0.989 |
| #4 | ∞ | 6.5" | 13.5" | -6.5" | 70040 | 835 | 0.552 |
| #5 | 13.5" | 4.25" | 4.25" | ∞ | ~1400000 | 864 | 0.937 |
| #8 | ∞ | 6.5" | ∞ | -4.25" | ~560490 | 724 | 0.869 |

Version #1 has the smallest $\gamma^2$ value, but because the first and last surfaces were changed, it has comparable validity with the versions #2 and #4. To confirm the choice, it is necessary to make the similar "re-polishing" as before for Dollond's doublet.
For "re-polished" version #1 the new radii, which save achromaticity, will be
$R_1 \approx -32$" (812.8mm); $R_2 \approx 6.5$" (165.1mm); $R_3 \approx 8.75$" (222.25mm); $R_4 \approx -6.25$" (158.75mm); and $\gamma^2$ drops down to 3600.
The new radii will be
for version #2: $R_1 \approx -32$"; $R_2 = R_3 \approx 8.75$"; $R_4 \approx -9.75$" (247.65mm); and $\gamma^2 \approx 89180$;
for version #4: $R_1 \approx -32$"; $R_2 \approx 6.5$"; $R_3 \approx 13.5$" (342.9mm); $R_4 \approx -5$" (127mm); and $\gamma^2 \approx 112400$.
As we can see both versions made step in the wrong direction. Therefore, after this verification for the doublet we have to select version #1 as most validity between the approximation versions. It is the same choice as for Linnell's doublet, which was expected.

**Conclusions**

The twenty possible versions of the flint-forward achromatic doublets in "thin-lens" approximation were considered. A qualitative comparative analysis of the versions by values of residual (spherical and chromatic) aberrations was carried out for the optical systems with non-zero thickness of the lenses and gaps between them (if it was required). Herewith the radii of the lens surfaces had the same values which were obtained from the formulas for "thin-lens" approximation. This approach makes it possible to evaluate the quality of the particular approximation version and its applicability for making of the real achromatic doublets.
Six of the twenty versions which principally could be used in optical practice were selected. Among the six versions exists only one (#8) which was undoubtedly attributed to Chester Moor Hall according to the testimony of W. Eastland. The analysis showed that version #8 as well as #2 and #4 have a big enough residual aberrations. These approximation versions have a mediocre quality relative to the versions #1, #3 and #5. They can be used at relative apertures slower than 1/23.
In accordance with introduced $\gamma^2$ criterion Linnell's and Ayscough's doublets were made by applying the approximation version #1. This conclusion from the article [1] was confirmed. However other conclusion relative to Dollond's approximation should be avoided. Version #2 which was suggested as that applied by Dollond has less validity (at least for Dollond's flint-forward doublet from the article) and it has worse quality of the initial approximation in comparison with version #3.
The versions #1 and #3 are comparable in quality approximation. Both can be applied at relative apertures slower than 1/15.
Ayscough and Dollond both decided to apply the additional correction of the flat surface on the flint lens. Most likely, the corrections were done to eliminate the deformation of the flat surface due to partial removal of internal (residual) stresses and Twyman effect, which take place every time after manufacturing a concave (or convex) surface on the opposite side. These changes in the initial approximations can be easily calculated before manufacturing stage, therefore the trial and error method was not required. Presumably the main portion of a manufacturing labor hours of the achromatic doublets the XVIII century opticians were spent on measuring the glass parameters with the required accuracy and on keeping the calculated radii to given tolerances.

**Acknowledgements**
Discussions about early telescopes with Yuri Petrunin was helpful for raising interest in the subject and it is much appreciated. I wish to also thank Sergey Petrunin for reading this article and correcting my grammar blunders.

## Appendix I

As in the article [2] the "thin-lens" approximation was used. In this case the total optical power $P$ of the doublet is calculated by next simple formula:

$$P \equiv \frac{1}{F} = P_f + P_c \qquad \text{(I.1)}$$

Here $P_f$; $P_c$ – optical powers of the flint and crown lenses respectively, at this $P_f < 0$; $P_c > 0$ and $P > 0$.
$F$ – the selected focal length of the achromatic doublet.
According to the used "thin-lens" approximation the optical powers for each lenses are found from formulas:

$$P_f = (n_f - 1)\left[\frac{1}{R_1} - \frac{1}{R_2}\right]$$

$$P_c = (n_c - 1)\left[\frac{1}{R_3} - \frac{1}{R_4}\right] \qquad \text{(I. 2)}$$

Here $n_f$; $n_c$ – refractive indexes of flint and crown glasses respectively at green-yellow wavelength.
The doublet will be achromatic if the optical powers will satisfy to next proportion:

$$\frac{P_c}{P_f} = -\frac{\nu_c}{\nu_f} \qquad \text{(I.3)}$$

Here $\nu_f$; $\nu_c$ – Abbe numbers of the used glasses. The negative sign is required, because the Abbe numbers are positive values, and always $\nu_c > \nu_f$
The four radii in (I.2) should be selected like that to satisfy two conditions:
That the doublet lens has focal length equal to the selected one in (I.1) and
That the achromatic doublet (I.3) eventually should be implemented.
Thus remain two free radii and they should be set in volitional way, for instance, to be equal to each other or to be flat or with any other preselected radius. So eleven versions are possible, which are shown on Figure 4 and the additional nine versions are shown on Figure I.
For selected pair of the glass, that means the parameters $n_f$; $\nu_f$; $n_c$; $\nu_c$ are known, the radii for flint and crown lenses for the different versions of the achromatic doublets can be calculated with help the formulas from Table I.

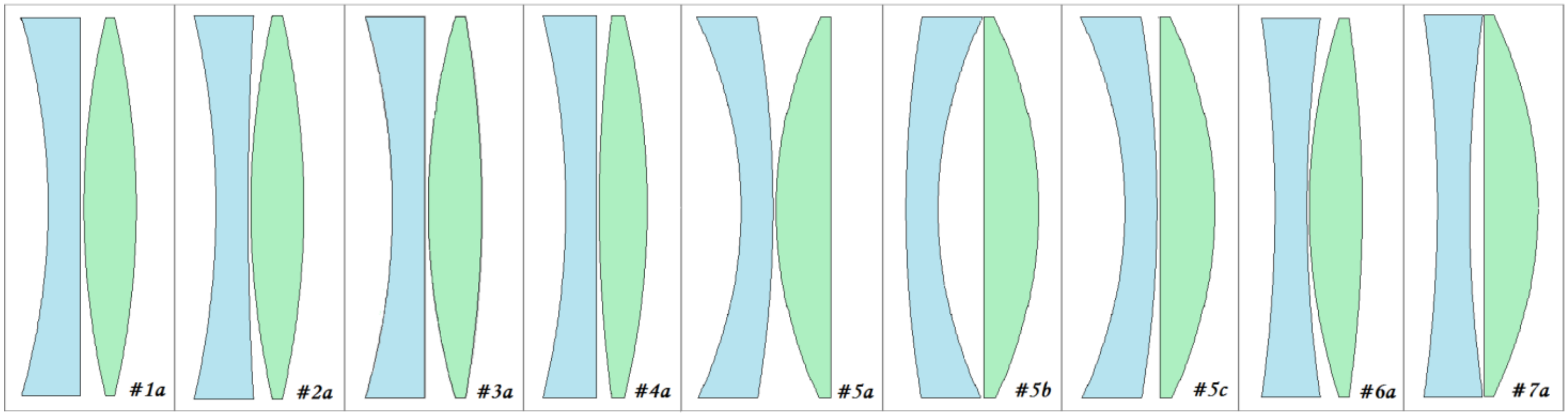

Figure I. The additional versions of the flint-forward achromatic doublets are shown. Versions #1a, #2a, #3a, 4a and #5a are different from versions #1, #2, #3, #4 and #5 respectively by reverse orientation of the flint lens (blue part). The versions #5b, #6a and #7a are different by reverse orientation of the crown lens (green part) for the corresponding numbers of the versions. In version #5c orientations for both lenses are changed relative to version #5. At all these changes the radii of lens surfaces are the same as at the original versions with corresponding numbers.

The versions from the additional list all have a wavefront errors from 0.9λ to 2.5λ, therefore are not of interest (also as the versions #6, #7, #9, #10, #11 foregoing). By this reason they were not included in Table 4, no necessity to overload the table also by these data.
The rows in the Table I are sometimes divided by the double lines. It means between them the radius values should be calculated by the same formulas.

Table I. The formulas for calculation of the radii for the flint-forward achromatic doublets.

| | $R_1$ | $R_2$ | $R_3$ | $R_4$ |
|---|---|---|---|---|
| #1 | ∞ | $\frac{1}{2}\frac{n_f-1}{n_c-1}\frac{\nu_c}{\nu_f}\cdot R$ | $R=2(n_c-1)\left[1-\frac{\nu_f}{\nu_c}\right]\cdot F$ | −R |
| #1a | $-R_2$ | ∞ | R | −R |
| #2 | $\frac{R}{1-2\frac{n_c-1}{n_f-1}\frac{\nu_f}{\nu_c}}$ | R | $R=2(n_c-1)\left[1-\frac{\nu_f}{\nu_c}\right]\cdot F$ | −R |
| #2a | −R | $-R_1$ | R | −R |
| #3 | ∞ | $R=(n_f-1)\left[\frac{\nu_c}{\nu_f}-1\right]\cdot F$ | R | $\frac{-R}{\frac{n_f-1}{n_c-1}\frac{\nu_c}{\nu_f}-1}$ |
| #3a | −R | ∞ | R | $R_4$ |
| #4 | ∞ | $R=(n_f-1)\left[\frac{\nu_c}{\nu_f}-1\right]\cdot F$ | $\frac{R}{\frac{n_f-1}{n_c-1}\frac{\nu_c}{\nu_f}-1}$ | −R |
| #4a | −R | ∞ | $R_3$ | −R |
| #5 | $\frac{R}{1-\frac{n_c-1}{n_f-1}\frac{\nu_f}{\nu_c}}$ | R | $R=(n_c-1)\left[1-\frac{\nu_f}{\nu_c}\right]\cdot F$ | ∞ |
| #5a | −R | $-R_1$ | R | ∞ |
| #5b | $R_1$ | R | ∞ | −R |
| #5c | −R | $-R_1$ | ∞ | −R |
| #6 | −R | $R=2\left(n_f-1\right)\left[\frac{\nu_c}{\nu_f}-1\right]\cdot F$ | R | $\frac{-R}{2\frac{n_f-1}{n_c-1}\frac{\nu_c}{\nu_f}-1}$ |
| #6a | −R | R | $R_4$ | −R |
| #7 | −R | $R=2\left(n_f-1\right)\left[\frac{\nu_c}{\nu_f}-1\right]\cdot F$ | $\frac{1}{2}\frac{n_c-1}{n_f-1}\frac{\nu_f}{\nu_c}\cdot R$ | ∞ |
| #7a | −R | R | ∞ | $-R_3$ |
| #8 | ∞ | $\left(n_f-1\right)\left[\frac{\nu_c}{\nu_f}-1\right]\cdot F$ | ∞ | $(n_c-1)\left[\frac{\nu_f}{\nu_c}-1\right]\cdot F$ |
| #9 | $\left(n_f-1\right)\left[1-\frac{\nu_c}{\nu_f}\right]\cdot F$ | ∞ | ∞ | $(n_c-1)\left[\frac{\nu_f}{\nu_c}-1\right]\cdot F$ |
| #10 | ∞ | $\left(n_f-1\right)\left[\frac{\nu_c}{\nu_f}-1\right]\cdot F$ | $(n_c-1)\left[1-\frac{\nu_f}{\nu_c}\right]\cdot F$ | ∞ |
| #11 | $\left(n_f-1\right)\left[1-\frac{\nu_c}{\nu_f}\right]\cdot F$ | ∞ | $(n_c-1)\left[1-\frac{\nu_f}{\nu_c}\right]\cdot F$ | ∞ |